\documentstyle[12pt]{article}
\newcommand{\CMP}[1]{{\em Commun. Math. Phys.} {\bf {#1}}}
\newcommand{\JMP}[1]{{\em J.~Math. Phys.} {\bf {#1}}}
\newcommand{\NP}[1]{{\em Nucl.Phys.~B} {\bf {#1}}}

\newcommand{\PL}[1]{{\em Phys. Lett.} {\bf {#1}}}
\newcommand{\PR}[2]{{\em Phys. Rev.} {#1} {\bf {#2}}}

\newcommand{\Map}{C_0^\infty}
\newcommand{\half}{\mbox{$\frac{1}{2}$}}

\newcommand{\SU}{{\rm SU}}

\newcommand{\gl}{{\rm gl}}

\newcommand{\eq}{\begin{equation}}
\newcommand{\eqend}{\end{equation}}
\newcommand{\eqa}{\begin{eqnarray}}
\newcommand{\nonueqa}{\begin{eqnarray*}}
\newcommand{\eqaend}{\end{eqnarray}}
\newcommand{\nonueqaend}{\end{eqnarray*}}
\newcommand{\nonu}{\nonumber \\ \nopagebreak}
\newcommand{\bma}[1]{\begin{array}{#1}}
\newcommand{\ema}{\end{array}}
\newcommand{\bc}{\begin{center}}
\newcommand{\ec}{\end{center}}

\newcommand{\Ref}[1]{(\ref{#1})}

\newcommand{\ii}{{\rm i}}

\newcommand{\om}{\omega}
\newcommand{\gam}{\gamma}
\newcommand{\Gam}{\Gamma}

\renewcommand{\phi}{\varphi}

\newcommand{\del}{\delta}

\newcommand{\tet}{\theta}

\newcommand{\eps}{\varepsilon}

\newcommand{\sign}{{\rm sign}}

\newfam\msbfam
\batchmode\font\twelvemsb=msbm10 scaled\magstep1 \errorstopmode
\ifx\twelvemsb\nullfont\def\Bbb{\bf}
\else	\catcode`\@=11
	\font\tenmsb=msbm10 \font\sevenmsb=msbm7 \font\fivemsb=msbm5
	\textfont\msbfam=\tenmsb
	\scriptfont\msbfam=\sevenmsb \scriptscriptfont\msbfam=\fivemsb
	\def\Bbb{\relax\ifmmode\expandafter\Bbb@\else
 		\expandafter\nonmatherr@\expandafter\Bbb\fi}
	\def\Bbb@#1{{\Bbb@@{#1}}}
	\def\Bbb@@#1{\fam\msbfam\relax#1}
	\catcode`\@=\active
\fi
\newcommand{\R}{{\Bbb R}}
\newcommand{\C}{{\Bbb C}}

\newcommand{\N}{{\Bbb N}}

\newcommand{\f}{\frac}
\newcommand{\cA}{{\cal A}}
\newcommand{\cB}{{\cal B}}
\newcommand{\cC}{\Omega}
\newcommand{\cD}{{\cal D}}
\newcommand{\cL}{{\cal L}}

\newcommand{\cH}{{\cal H}}
\newcommand{\cF}{{\cal F}}

\newcommand{\cG}{{\cal G}}

\newcommand{\cS}{{\cal S}}
\newcommand{\cV}{{\cal V}}
\newcommand{\cI}{{\cal J}}

\newcommand{\ccr}[2]{{[} {#1},{#2} {]} }        
\newcommand{\car}[2]{{\{} {#1},{#2} {\}} }        

\newcommand{\Tra}[1]{{\rm Tr} \left({#1}\right)}          
\newcommand{\TraC}[1]{{\rm Tr}_C \left({#1}\right)}          
\newcommand{\trac}[1]{{\rm tr}_{N} \left({#1}\right)}          
\newcommand{\trc}{{\rm tr}_{N}}          

\newcommand{\dd}{{\rm d}}
\newcommand{\D}{D\!\!\!\!\slash}

\newcounter{saveeqn}
\newcounter{App} 

\newcommand{\alpheqn}{%
\stepcounter{equation}
\setcounter{saveeqn}{\value{equation}}%
\setcounter{equation}{0}%
\renewcommand{\theequation}{\arabic{saveeqn}\alph{equation}}}
\newcommand{\reseteqn}{\setcounter{equation}{\value{saveeqn}}%
\renewcommand{\theequation}{\arabic{equation}}}

\begin{document}

\begin{flushright}
{August 1, 1995}
\end{flushright}
\vspace{.4cm}
\renewcommand{\thefootnote}{\alph{footnote}}

\begin{center}
{\Large \bf Descent equations of Yang--Mills anomalies
in noncommutative geometry}\\
\vspace{1 cm}
{\large Edwin Langmann}\\
\vspace{0.3 cm}
{\em Theoretical Physics, Royal Institute of Technology, S-10044 
Stockholm, Sweden}\\
\end{center}

\setcounter{footnote}{0}
\renewcommand{\thefootnote}{\arabic{footnote}}

\newcommand{\CC}{\cC}
\newcommand{\Cn}{\hat\CC_d}
\newcommand{\hd}{\hat{\dd}}
\newcommand{\hint}{\hat{\int}}
\newcommand{\map}{Map}
\newcommand{\dRn}{\CC_{{\rm dR},d}}
\newcommand{\dR}[1]{\CC_{{\rm dR},{#1}}}
\newcommand{\CCn}{\hat\CC_{{\rm dR},d}}
\newcommand{\CCs}[1]{\hat\CC_{{\rm dR},{#1}}}
\newcommand{\SS}{\cS}
\newcommand{\DD}{\cD}
\newcommand{\Sn}{\hat{\cS}_{{\rm dR},d}}
\newcommand{\tom}{\tilde\om}
\newcommand{\hdel}{\hat\delta}
\newcommand{\dt}{\partial_t}
\renewcommand{\kill}[1]{{#1}\!\!\!\!\!\slash\,\,}
\newcommand{\ccF}{\tilde\cF}
\newcommand{\dDel}{\tilde\Delta}
\newcommand{\cXi}{\tilde\Xi}
\renewcommand{\iota}{\cI}
\newcommand{\hI}{\hat I}

\begin{abstract}
Consistent Yang--Mills anomalies $\int\om_{2n-k}^{k-1}$
($n\in\N$, $ k=1,2, \ldots ,2n$) as described collectively by
Zumino's descent equations
$\delta\om_{2n-k}^{k-1}+\dd\om_{2n-k-1}^{k}=0$ starting with the
Chern character $Ch_{2n}=\dd\om_{2n-1}^{0}$ of a principal $\SU(N)$
bundle over a $2n$ dimensional manifold are considered (i.e.\
$\int\om_{2n-k}^{k-1}$ are the Chern--Simons terms ($k=1$), axial
anomalies ($k=2$), Schwinger terms ($k=3$) etc.\ in $(2n-k)$
dimensions). A generalization in the spirit of Connes' noncommutative
geometry using a minimum of data is found. For an arbitrary
graded differential algebra $\CC=\bigoplus_{k=0}^\infty \CC^{(k)}$
with exterior differentiation $\dd$, form valued functions $Ch_{2n}:
\CC^{(1)}\to \CC^{(2n)}$ and $\om_{2n-k}^{k-1}:
\underbrace{\CC^{(0)}\times\cdots \times \CC^{(0)}}_{\mbox{{\small $(k-1)$
 times}}} \times \CC^{(1)}\to \CC^{(2n-k)}$ are constructed which are
connected by generalized descent equations
$\delta\om_{2n-k}^{k-1}+\dd\om_{2n-k-1}^{k}=(\cdots)$.
Here $Ch_{2n}= (F_A)^n$ where $F_A=\dd(A)+A^2$ for $A\in\CC^{(1)}$,
and $(\cdots)$ is not zero but a sum of graded commutators which
vanish under integrations (traces).
The problem of constructing Yang--Mills anomalies on a given graded
differential algebra is thereby reduced to finding an interesting
integration $\int$ on it. Examples for graded differential algebras
with such integrations are given and thereby noncommutative
generalizations of Yang--Mills anomalies are found.
\end{abstract}

\newpage

\section{Introduction}

Connes' noncommutative geometry (NCG) is a very influential subject in 
mathematics unifying notions and ideas from geometry and functional 
analysis \cite{C1,C2} (for review see \cite{Crev}).  Since recently it is 
also receiving broad attention from the physics community.  This was 
triggered by a new geometric interpretation of the structure of the 
standard model including Higgs sector on the classical level based on NCG 
\cite{ConnesLott} (see also \cite{Crev} and references therein).  Even more 
interesting, however, is the possibility that NCG could provide a natural 
mathematical framework for {\em quantum} field theory, considering that a 
deeper mathematical understanding of regularization and renormalization of 
3+1 dimensional quantum gauge theories like QCD has been a major challenge 
in theoretical physics up to this day.  Progress in this direction should 
not only lead to a better understanding of the nonperterburbative structure 
of these fundamental theories of nature but also provide new calculation 
tools for analyzing them.

One hint that NCG could be useful in this context is that one of the
most widely used regularization schemes for quantum gauge theories
--- dimensional regularization \cite{tH} --- is done by formally extending the
models to $(4-\epsilon)$--dimensional spacetime, and
that NCG provides the means to naturally extend the
setting of Yang--Mills gauge theories --- connections on a principal
$\SU(N)$--bundle over a manifold $M$ --- to more general situations
without underlying manifold. In fact, all that is required is a
graded differential algebra (GDA) $(\CC,\dd)$ which is a differential
complex, $\CC=\bigoplus_{k=0}^\infty\CC^{(k)}$ with a differentiation
$\dd:\CC\to\CC$  mapping $\CC^{(k)}$ to $\CC^{(k+1)}$ such that
$\dd^2=0$, carrying a compatible algebra structure (for precise
definitions see next Section). Then elements $A\in\CC^{(1)}$ can be
interpreted as generalized Yang--Mills connections, and
$X\in\CC^{(0)}$ as infinitesimal gauge transformations acting on
$\CC^{(1)}$ by $A\to \dd(X)+\ccr{A}{X}$. To formulate also the
Yang--Mills action one needs in addition an integration $\int$ and a
Hodge--$*$ operation on $\CC$. In case of ordinary Yang--Mills theory,
this GDA is, of course, given by the de Rham
forms on the spacetime  (or space) manifold $M$, but there are
interesting examples for GDA $(\CC,\dd)$ based on algebras
of Hilbert space operators and which naturally generalize the de Rham
forms \cite{C1,C2} (see also next Section).

One important aspect of quantized gauge theories is the occurrence of 
anomalies and other topological terms like axial anomalies, Schwinger 
terms, or Chern characters (for review see e.g.\ \cite{Jackiw}).  An 
essential requirement to any regularization procedure used in this context 
is that it correctly generates the anomalies.  This is nontrivial --- we 
recall the well--known difficulties to derive, or even formulate, anomalies 
(topological terms) in lattice gauge theories (see e.g.\ \cite{Luscher}).  
The motivation for the present work was an example where a generalized 
Yang--Mills setting $(\CC,\dd)$, as discussed above, has been used 
successfully for regularizing certain fermion Yang--Mills systems and 
giving a mathematically rigorous quantum field theory derivation of 
Schwinger terms \cite{MR,M,L,LM1} (for a more leisurely discussion of the 
following see \cite{L2}).  One basic idea here is to consider not only the 
gauge group $\cG$ and the set of Yang--Mills fields $\cA$, but certain 
algebras of Hilbert space operators in which $\cG$ and $\cA$ can be 
naturally embedded.  The defining relations of these algebras are Schatten 
ideal conditions which can be regarded as abstract characterizations of the 
degree of ultraviolet divergences occurring in the spacetime dimension 
under consideration.  Interestingly, this leads exactly to a GDA 
$(\CC,\dd)$ which plays a fundamental role in NCG \cite{C1}.  One is 
thereby naturally lead to a universal Yang--Mills setting based on such GDA 
of Hilbert space operators (for another application of this in quantum 
field theory see \cite{R,FR}).  One advantage of using this extended 
setting for regularization is that it forces one to concentrate on 
essentials.  This leads to very transparent, despite more general, 
constructions.  One remarkable result is an universal Schwinger term in 
(3+1) dimensions
\cite{MR} generalizing the Schwinger term occurring in the Gauss' law
commutators in chiral QCD(3+1) \cite{LM1}. This former Schwinger terms has
a natural interpretation as generalization of the latter to NCG if one
regards the (conditional) Hilbert space trace as generalization of
integration of de Rham forms \cite{LM1,L1}. Having this one example where
NCG appears in a quantum field theory derivation of anomalies, it is
natural to ask whether also other anomalies have similar generalizations to
the universal Yang--Mills setting, or, more generally, to other GDA.
The structure of these generalized anomalies can be expected to directly
hint to how they arise from regularizing a quantum gauge theory. For
example, the regularization procedure used in \cite{L} to derive the
anomalous Gauss' law commutators in chiral QCD trivially generalizes from
(3+1) to higher dimensions once the higher dimensional universal Schwinger
terms are known.

In this paper we give a positive answer to this question treating all kinds 
of anomalies and arbitrary GDA at once.  We show that certain `raw data' 
for all anomalies and topological terms can be constructed in {\em any} GDA 
$(\CC,\dd)$, and these give rise to anomalies (nontrivial cocycles) if and 
only if there is an appropriate nontrivial integration on $(\CC,\dd)$.  
Since the GDA associated with the universal Yang--Mills setting have such 
nontrivial integrations (the conditional Hilbert space trace), we obtain 
the universal generalizations of all anomalies as a special case 
(including, of course, all higher dimensional universal Schwinger terms).

We believe that many (perhaps all) interesting gauge invariant 
regularization schemes for Yang--Mills systems can be associated with some 
GDA as described in this paper.  Thus our result should be fundamental for 
quantum gauge theories.  It also provides an explanation how anomalies with 
their rich differential geometric structure can arise from explicit field 
theory calculations.  The latter are done by calculating Feynman diagrams 
(see \cite{Jackiw} and references therein) which can be interpreted as 
(regularized) traces of certain Hilbert space operators \cite{LM2}; our 
result shows how the differential geometric structure of anomalies can be 
present already on the level of Hilbert space operators.  We feel that this 
adds strong support to the expectation that NCG is relevant for quantum 
gauge theories.

To be more specific, we recall Zumino's descent equations
$\delta\bar\omega^{k-1}_{2n-k}+ \dd \bar\omega^{k}_{2n-k-1}=0$
\cite{Zumino} providing a collection of intriguing relations
between all Yang--Mills anomalies (the reason for the
bar over $\om$ will become clear immediately). These relations imply that
the integrals $\int_{M}\bar\omega^{k-1}_{2n-k}$  over $(2n-k)$ dimensional
manifolds $M$ without boundary are $(k-1)$--cocycles, and these are exactly
the Chern Simons terms $(k=1)$, axial anomalies $(k=2)$, Schwinger
terms $(k=3)$ etc., occurring in $(2n-k)$ dimensional gauge
theories.  Our result is a generalization of Zumino's descent
equations to all GDA $(\CC,\dd)$. For all
$X_i\in\CC^{(0)}$ and $A\in\CC^{(1)}$ we construct forms
$\om^{k-1}_{2n-k}(X_1,\ldots,X_{k-1};A)\in\CC^{(2n-k)}$ which are
multilinear and antisymmetric in the $X_i$ and obey the generalized
descent equations
\eq
\label{desc}
\delta \omega^{k-1}_{2n-k} + \dd\omega^{k}_{2n-k-1} = (\cdots)
\quad k=1,2,\ldots, 2n, \quad (n\in\N) \eqend
(for definition of $\del$ see \Ref{del}, for full equations 
(\ref{desca}--e)) where $(\cdots)$ is not zero but a sum of graded 
commutators which vanish under integrations $\int$.  It is interesting to 
note that these descent equations generalize Zumino's even for the case of 
de Rham forms: we recall that the forms mentioned above are traces of 
matrix valued forms, $\bar\omega^{k-1}_{2n-k}=\trac{\omega^{k-1}_{2n-k}}$, 
and our equations are without the matrix trace $\trc$ (the terms $(\cdots)$ 
vanish under $\trc$).  Our generalization is natural since $\trc$, 
which we regard as partial integration, is not available for general GDA.

As mentioned, these forms $\omega^{k-1}_{2n-k}$ are raw
data from which the anomalies are obtained by integration $\int$; the
descent equations and basic properties of integration guarantee that
they obey the appropriate consistency conditions
$\del\int\omega^{k-1}_{2n-k}=0$ (cocycle relations).
Thus our result reduces the question of existence of anomalies for
general GDA $(\CC,\dd)$ to existence of a nontrivial integration
$\int$ on it. We stress that this latter question is still very
nontrivial. In fact, our result makes explicit that all
topological information is contained in such an
integration: the raw data for topological terms
$\omega^{k-1}_{2n-k}$ {\em always} exist and give rise to
$(k-1)$--cocycles by integration $\int$; weather these are nontrivial
solely depends on the integration $\int$ and has to be checked for
each case individually.

We note that there are other generalizations of Zumino's descent equations
in the literature \cite{CM,Stora}.  These generalize the equation for
$\bar\omega^{k-1}_{2n-k}$ and therefore also need the notation of
(partial) integration.  Our result is more general since it does not
require an integration.  In fact, we believe that it is the most general
result in this direction and very much in the spirit of NCG since it uses a
minimum amount of input data and strips the descent equations from a
differential geometric setting to the bones of simple algebraic relations.
(Note also that we do not have a graded double complex since
$\delta\dd+\dd\delta$ is not zero on forms $\omega^{k-1}_{2n-k}$ but only
on $\bar\omega^{k-1}_{2n-k}$!) Also, our proof is very simple and
direct --- we give an explicit formula for the forms
$\omega^{k-1}_{2n-k}$ and prove by explicit calculation that our
generalized descent equations are obeyed.  We did not attempt to find a
geometric interpretation of these equations \Ref{desc}, but since the
nontrivial r.h.s.  $(\cdots)$ have a very simple, suggestive form (cf.\
(\ref{desca}--e)), we believe that such an interpretation is possible and
should be interesting.

There are two important examples for GDA of bounded Hilbert space operators
which generalize the de Rham forms to NCG.  In one differentiation is given
by the graded commutator with the Dirac operator and integration by Dixmier
trace \cite{C2}, in the other differentiation is graded commutation with
the {\em sign} of the Dirac operator and integration is the Hilbert space
trace \cite{C1} (see also \cite{Crev}).  In this paper we use only the
latter as example since it is this GDA which naturally appears in quantum
field theory.  Also, the noncommutative generalization of anomalies to the
former is discussed in \cite{CM}.

To avoid confusion, we already point out here that we use Grassmann
numbers $\tet_i$ for writing down the formulas for the forms
$\omega^{k-1}_{2n-1}$, but this is only a matter of convenience and one
could do without at the cost of having longer expressions and proofs.
However, these Grassmann numbers are a convenient means to keep track of the
combinatorics and order of terms --- they are a substitute for the
symmetrized trace used by Zumino \cite{Zumino} which is not available in
our general setting.

The plan of this paper is as follows.  To make the paper self--contained 
and have nontrivial examples to which our result can be applied, we 
summarize the basis definitions and examples of GDA and integrations in 
Section 2.  This Section is also meant as an introduction to basic notions 
from NCG \cite{Crev} relevant for quantum field theory.  Of mathematical 
interest might be our definition of partial integration on GDA 
(generalizing integration of de Rham forms over manifolds with boundary) 
which seems somewhat different from the one in
\cite{Crev}.  Our main result and its proof are in Section 3.  In the final
Section we discuss some interesting special cases including the universal
anomalies mentioned.

{\bf Notation:} We denote as $\N_0$ the non--negative integers, $\C$ the
complex numbers, $\gl_N$ the complex algebra of $N\times N$ matrices
with algebra product the usual matrix multiplication,
$\ccr{a}{b}_\pm=ab\pm ba$, $\ccr{a}{b}=ab - ba$ and
$\car{a}{b}=ab + ba$.

\section{Prerequisites}

\subsection{Graded differential algebra (GDA)} A GDA $(\CC,\dd)$ is a
$\N_0$--graded associative algebra over the $\C$,
$\CC=\bigoplus_{k=0}^\infty \CC^{(k)}$, with a linear map $\dd:
\CC\to\CC$ such that $\dd^2=0$, and for all
$\om_{k,\ell}\in\CC^{(k,\ell)}$, $k,\ell\in\N_0$,
\eq
\label{GDA}
\om_k \om_\ell\in\CC^{(k+\ell)},\quad \dd \om_k\in\CC^{(k+1)},\quad \dd(\om_k
\om_\ell)= \dd(\om_k) \om_\ell + (-)^k \om_k \dd(\om_\ell). \eqend
(we write the algebra product in $\CC$ as $(\om,\om')\to \om\om'$).

We denote elements $\om_k\in\CC^{(k)}$ as $k$--forms and $\dd$ as
exterior differentiation. Note that by definition,
all linear combinations of elements $u_0\dd u_1\cdots \dd u_k$
with $u_i\in\CC^{(0)}$ are $k$--forms.

{\bf Example 1:} The simplest example for a GDA is the de Rham GDA
$(\dRn,\dd)$ of de Rham forms on $\R^d$ which we define as follows:
$\dRn^{(0)}=\Map(\R^d;\gl_N)$ is the algebra of $\gl_N$--valued
$C_0^\infty$ functions on $\R^d$, and for $k\geq 1$, $\dRn^{(k)}$ is the
vector space generated by elements $u_0\dd u_1\cdots \dd u_n$,
$u_i\in\dRn^{(0)}$, where $\dd$ is the usual exterior differentiation of de
Rham forms.

{\bf Example 2:}
Let $\cH$ be a separable Hilbert space. We denote as $\cB$ and
$\cB_1$ the bounded and trace class operators on $\cH$, respectively,
and $\cB_p=\{a\in \cB|\, (a^*a)^{p/2}\in \cB_1\}$ for $p>0$ are the
Schatten classes. We recall that $a\in \cB_{\f{n}{p}}$,
$b\in \cB_{\f{n}{q}}$ and $n,p,q >0$ implies
$ab\in \cB_{\f{n}{p+q}}$ \cite{Simon}. Let $\eps\in \cB$ be a grading
operator, i.e.\ $\eps=\eps^*=\eps^{-1}$. We will use the
notation
\eq
a=a_+ + a_-,\quad a_\pm = \half\left(a \pm \eps a
\eps\right )\quad \forall a\in \cB
\eqend
so that $(ab)_+=a_+b_+ + a_-b_-$, $(ab)_- = a_+b_- +a_-b_+$.

Let $d\in \N$.  We define
\alpheqn
\eq
p=\half(d+1)
\eqend
and
\eqa
\Cn^{(0)}= \left\{ a\in\cB\left|\, a_-\in\cB_{2p} \right.\right\} \nonu
\Cn^{(2k-1)}= \left\{ a\in\cB\left|\, a_+\in\cB_{\f{{2p}}{2k}},
a_-\in\cB_{\f{{2p}}{2k-1}} \right.\right\}\nonu
\Cn^{(2k)}= \left\{ a\in\cB\left|\, a_+\in\cB_{\f{{2p}}{2k}},
a_-\in\cB_{\f{{2p}}{2k+1}} \right.\right\}\nonu
k=1,2,\ldots.
\eqaend
Then
\eq
\hd \om_k\equiv \ii\left(\eps \om_k - (-)^k \om_k\eps \right)\quad
\forall \om_k\in\Cn^{(k)}
\eqend
\reseteqn
defines an exterior differentiation $\hd$ on
$\Cn=\bigoplus_{k=0}^\infty\Cn^{(k)}$ which makes
$(\Cn,\hd)$ into a GDA.

Note that the algebra multiplication in this GDA is equal to the operator 
product in $\cB$ (the latter follows from $\eps^2=1$ which implies 
$\eps\ccr{\eps}{a}\eps=-\ccr{\eps}{a}$ for all $a\in\cB$).

For $d=2n$ {\em even}, we introduce
another  grading operator $\Gam$ on $\cH$ which
anticommutes with $\eps$, $\eps\Gam = - \Gam\eps$.
For the introduction of integrations below we
need to restrict $(\Cn,\hd)$ to a subalgebra $(\CCn,\hd)$
defined as $\CCn = \bigoplus_{k=0}^\infty \CCn^{(k)}$,
\eq
\label{GC}
\CCn^{(k)} = \left\{\om_k\in\Cn^{(k)} \left|\,
\om_k\Gam -(-)^k\Gam \om_k = 0 \right.\right\}\quad (d=2n)
\eqend
(it is easy to see that this indeed defines a GDA).
To simplify notation we write $\Gam^{d-1}=\Gam$
and $1$ (identity) for $d$ even and odd,
respectively, and for $d$ odd, $\CCn=\Cn$.

The above mentioned  embedding of the de Rham GDA $(\dRn,\dd)$
in $(\CCn,\hd)$ is as follows. Consider the Hilbert space
\eq
\label{cH}
\cH=L^2(\R^d)\otimes \C^\nu_{spin}\otimes\C^N
\eqend
where
$\nu=2^{[d/2]}$ can be interpreted as the number of spin indices
($[d/2]=(d-1)/2$ for $d$ odd and $[d/2]=d/2$ for $d$ even).
With $\gam_i$ the usual selfadjoint $\gam$--matrices
acting on $\C^\nu_{spin}$ and obeying
$\gam_i\gam_j+\gam_j\gam_i=2\delta_{ij}$ (for explicit
formulas see e.g.\ \cite{L1}), the free Dirac operator
$\D_{\, 0}=\sum_{i=1}^d \gam_i(-\ii)\f{\partial}{\partial x_i}$
on $\R^d$ defines a selfadjoint operator on $\cH$ which we also
denote as $\D_{\, 0}$. For $d=2n$ even
we have an additional $\gam$--matrix $\gam_{d+1}$ which defines a
grading operator $\Gam$ on $\cH$.

We recall that every $X\in\Map(\R^d,\gl_N)$ can be naturally identified
with a bounded operator $\hat X$ on $\cH$, $(\hat Xf)(x)=X(x)f(x)$ for all
$f\in \cH$ ($x\in\R^d$).

We define $\eps = \sign(\D_{\, 0})$ where $\sign(x) = +1 (-1)$ for
$x\geq 0$ ($x<0$) (using the spectral theorem for self--adjoint
operators \cite{RS1}). For $d=2n$ even, $\Gam=\gam_{d+1}$ anticommutes with
$\eps$, and it obviously commutes with all $\hat X$ for
$X\in\Map(\R^d,\gl_N)$.

One can prove
\alpheqn
\eq
\label{embed}
\mbox{$X\in \Map(\R^d,\gl_N)$
$\Rightarrow$  $\hat X\in \CCn^{(0)}$}
\eqend
(see e.g.\ \cite{MR}) implying that
\eq
X_0\dd X_1\cdots X_k \mapsto (\ii)^k
\hat X_0\ccr{\eps}{\hat X_1}\cdots\ccr{\eps}{\hat X_k}\quad
\forall X_i\in\dRn^{(k)}
\eqend
\reseteqn
provides an embedding $\om\mapsto \hat \om$ of $(\dRn,\dd)$ in
$(\CCn,\hd)$

We shall refer to $(\CCn,\hd)$ as {\em universal de Rham GDA}.

\subsection{Integrations} A GDA with integration is a triple
$(\CC,\dd,\int)$ such that $(\CC,\dd)$ is a GDA, and there is a linear map
$\int:\CC \to \C$ obeying
\eq
\label{int}
\int\dd \om_k = 0,\quad \int \om_k \om_\ell = (-)^{k\ell}\int \om_\ell \om_k
\quad \forall \om_{k,\ell}\in\CC^{(k,\ell)}.
\eqend
(The first of these relations is Stokes' theorem and the second graded
commutativity of forms under the integration.)
If $\int$ in nonzero only on $\CC^{(d)}$ for some
$d\in\N_0$ we say that $\CC$ has dimension $d$ and indicate this by
writing $\CC=\CC_d$.

{\em Remark:} A GDA with integration is called {\em cycle} in \cite{Crev}.

{\bf Example 1:} The de Rham GDA $(\dRn,\dd)$
has the natural integration defined as $\int \om_k=0$ for
$\om_k\in\dRn^{(k\neq d)}$ and
\eq
\int \om_d= \int_{\R^d}\trac{\om_d}\quad \forall \om_d\in\dRn^{(d)}
\eqend
(integration of de Rham forms as usual; $\trac{\cdot}$ is the
trace of $N\times N$ matrices).

{\bf Example 2:} We define the conditional trace as
\eq
\label{trac}
\TraC{a}\equiv \Tra{a_+} \quad \mbox{for all $a\in\cB$ with
$a_+\in\cB_1$}
\eqend
where $\TraC{\cdot}$ is the usual Hilbert space trace on $\cH$. Then
$\hint \om_k=0$ for all $\om_k\in\CCn^{(k\neq d)}$ and
\alpheqn
\eq
\label{hint}
\hint \om_d = \f{1}{c_d}\TraC{\Gam^{d-1}\om_d}
\quad \forall \om_d\in\CCn^{(d)},
\eqend
where
\eq
\label{cd}
c_d = (2\ii)^{[d/2]} \f{1}{d(2\pi)^d}
\f{2\pi^{d/2}}{\Gamma(d/2)} 
\eqend
\reseteqn
is a convenient normalization constant (explained below),  defines an
integration on $(\CCn,\dd)$. Indeed, for $\om_d\in\CCn^{(d)}$,
$$
(\Gam^{d-1}\om_d)_+ = \left\{ \bma{cc} (\om_d)_+ & \mbox{ for $d$ odd}\\
\Gam (\om_d)_- &\mbox{ for $d$ even} \ema\right.
$$
which is $\cB_1$ by definition. The relations \Ref{int} can be easily
checked. (The first relation of \Ref{int} (Stokes' theorem) is
trivial since
$$
(\Gam^{d-1} \hd \om_{d-1})_+ = -\mbox{$\f{\ii}{2}$}
\Gam^{d-1}\eps \hd^2(\om_{d-1}) =0
$$
The second follows from $$
\left(\Gam^{d-1} (\om_k \om_{d-k} - (-)^{k(d-k)}\om_{d-k}\om_k) \right)_+ =
\ccr{(\Gam^{d-1} \om_k)_+}{(\om_{d-k})_+} + \ccr{(\Gam^{d-1}
\om_k)_-}{(\om_{d-k})_-}
$$
and cyclicity of trace (recall that $\Tra{\ccr{a}{b}}=0$ if at least
one of $a,b\in\cB$ is compact and $ab$ and $ba$ are $\cB_1$).
Note that in case $d$ is even, we used that $\Gam \om_k = (-)^k
\om_k\Gam$; this is why we had to restrict ourselves from $\Cn$ to
$\CCn$.

The embedding $\om\mapsto\hat \om$ of de Rham GDA $(\dRn,\dd)$
in $(\CCn,\hd)$ as discussed above naturally extends to these
integrations. One has
\eqa
\label{result}
\ii^d\TraC{\Gam^{d-1}
\hat X_0\ccr{\eps}{\hat X_1}\cdots \ccr{\eps}{\hat X_d}} =
c_d \int\trac{X_0 \dd X_1\cdots \dd X_d} \nonu \forall
X_i\in\Map(\R^d;\gl_N)
\eqaend
(for an elementary proof see \cite{L1})
which shows that $\hint\hat \om=\int \om$ for all $\om\in\dRn$.

\subsection{Partial Integrations}
A GDA with partial integration structure (GDAPI) is a
quintuple $(\CC,\dd,\int,{^*}\CC,\partial)$ such that,
(i) $(\CC,\dd)$ is a GDA,
(ii) ${^*}\CC=\bigoplus_{k=0}^\infty
{^*}\CC^{(-k)}$ is graded cycle, i.e.\ a $\N_0$--graded vector
space with a linear map $\partial:{^*}\CC\to{^*}\CC$ such that
$\partial Q^{(-k)}\in {^*}\CC^{(-k-1)}$ for all
$Q^{(-k)}\in {^*}\CC^{(-k)}$ and $\partial^2=0$,
(iii) there is a bilinear map $\int:
{^*}\CC\times \CC \to \C, (Q,\om)\mapsto \int_Q \om$ such that
\eq
\label{GDAPI}
\int_{Q}\dd \om_k = \int_{\partial Q} \om_k,\quad
\int_{Q} \om_k \om_\ell = (-)^{k\ell}\int_{Q} \om_\ell \om_k
\eqend
for all $\om_{k,\ell}\in\CC^{(k,\ell)}$ and $Q\in {^*}\CC $.
(The first of these is the generalization of
Stokes' theorem for manifolds with boundaries.)
A GDAPI is of dimension $d$
if $\int$ is nonzero only on $\bigoplus_{k=0}^d
{^*}\CC^{(-k)}\times \CC^{(d-k)}$, and we write in this case
$\CC=\CC_d$ and ${^*}\CC={^*}\CC_d$.

In our examples below, GDAPI $(\SS,\dd,\int,{^*}\SS,\partial)$ arise
from a given GDA $(\CC,\dd,\int)$ with integration
as follows.
(i) $(\SS,\dd,\int)$ is a subcomplex of, i.e. a GDA contained in,
$(\CC,\dd,\int)$,
(ii) ${^*}\SS=\bigoplus_{k=0}^\infty {^*}\SS^{(-k)}$ is a graded complex
such that for each $Q\in{^*}\SS$ and $0<\epsilon < 1$ there
is a $\eta_\epsilon(Q)\in \CC$ such that
\alpheqn
\eq
\lim_{\epsilon\searrow 0}\int \eta_\epsilon(Q)\om 
\eqend
exist for all $\om\in\SS$ and, (iii)  for all
$\om_{k,\ell}\in\SS^{(k,\ell)}$,
\eq
\label{ntr}
\lim_{\epsilon\searrow 0}\int \eta_\epsilon(Q)\left(\om_{k}
\om_\ell -(-)^{k\ell}\om_\ell \om_k\right)= 0.
\eqend
\reseteqn
We then set
\eq
\int_{Q}\om\equiv
\lim_{\epsilon\searrow 0}\int \eta_\epsilon(Q)\om\nonu,
\eqend
and  $\partial Q$ is defined by
\eq
\eta_\epsilon(\partial Q)_{-k} \equiv (-)^{-k-1}\dd\eta_\epsilon(Q)_{-k}
\eqend
(we use the notation $\CC\ni\eta=\sum_k\eta_{-k}$ with
$\eta_{-k}\in\CC^{(-k)}$).
Stokes' theorem in \Ref{GDAPI} then trivially follows from
$\int\dd\left(\eta_\epsilon(Q)\om \right)=0$. Thus it is
\Ref{ntr} which is the nontrivial condition determining whether
$\SS$ and ${^*}\SS$ are compatible so as to form a GDAPI.

Obviously $(\SS,\dd,\int,{^*}\SS,\partial)$ is of dimension $d$ if
$(\CC,\dd,\int)$ is.

{\em Remark:} One can identify elements in ${^*}\SS$ with equivalence
classes $Q=\eta/\!\!\sim $ of maps
$\eta:(0,1)\to\CC,\epsilon\mapsto\eta_\epsilon$ such that
$\lim_{\epsilon \searrow 0}\int\eta_\epsilon \om$ exist for all
$\om\in\SS$, with the equivalence relation $$
\eta\sim\eta' \Leftrightarrow
\lim_{\epsilon \searrow 0}
\int\left(\eta_\epsilon-\eta_\epsilon'\right)\om=0\quad \forall \om\in\SS .
$$

{\bf Example 1:} Let $\DD_d^{(0)}$ be the set of $d$--dimensional
compact submanifolds $D$ of $\R^d$ with boundary $\partial D$ which
is a $(d-1)$--dimensional compact
$C^\infty$ manifold, and $\DD_d^{(-k)}$ be the set of all
$(d-k)$--dimensional compact $C^\infty$ manifolds of the form
$D^{(-k)}= D_0\cap\partial D_1\cap\cdots\cap\partial D_k$ with
$D_i\in\DD_d^{(0)}$. Then $(\dRn,\dd,\int,\DD,\partial)$ is a GDAPI
with $\partial$ the usual boundary operation and
$
\int_{D^{(-k)}}\om_{d-k} \equiv \int_{D^{(-k)}}\trac{\om_{d-k} }
$
the integration of an $(d-k)$-form over an $k$--dimensional
submanifold $D^{(-k)}$ of $\R^d$ as usual. With $\chi^{}_D(x)$ the
characteristic function of $D\in\DD_d^{(0)}$ ($\chi^{}_D(x)=1$ for
$x\in D$ and $=0$ otherwise) we can formally write
\nonueqa
\int_{D_0\cap\partial D_1\cap\cdots\cap\partial D_k}\trac{\om_{d-k}} =
\int_{\partial D_1\cap\cdots\cap\partial D_k}\trac{\chi^{}_{D_0}\om_{d-k}}
= \\
\int_{D_1\cap\partial D_2\cap\cdots\cap\partial D_k}
\trac{\dd\left(\chi^{}_{D_0}\om_{d-k}\right)} = \cdots \\
=\int_{\R^d}\trac{\chi^{}_{D_k}\dd\chi^{}_{D_{k-1}}\cdots \dd\chi^{}_{D_1}
\dd\left(\chi^{}_{D_0}\om_{d-k}\right)} \\
= (-)^{k}
\int_{\R^d}\trac{\dd \chi^{}_{D_k} \cdots \dd\chi^{}_{D_1}
\chi^{}_{D_0}\om_{d-k}}
\nonueqaend
where we used repeatedly $\partial^2=\dd^2=0$ and Stokes' theorem for
integration of de Rham forms over manifolds with boundaries.
Introducing approximate $\delta$--functions $\delta^{(\epsilon)}
\in C_0^\infty(\R^d;\R)$ for $0 < \epsilon < 1$ obeying
$\delta^{(\epsilon)}(x)\geq 0$, for all $x\in\R^d$, $=0$ for
$|x|\geq \epsilon$ and $\int_{\R^d}\dd^d x\,
\delta^{(\epsilon)}(x)=1$, we define
$\chi^{(\epsilon)}_{D}(x)=\int_{\R^d}\dd^d y\,
\delta^{(\epsilon)}(x-y)\chi^{}_D(y)$. With that, our formal
calculation above implies
\alpheqn
\eq
\int_{D^{(-k)}}\trac{\om_{d-k}}
= \lim_{\epsilon \searrow 0} \int_{\R^d}\trac{\eta_\epsilon(D^{(-k)})
\om_{d-k}}
\eqend
with  \eq
\eta_\epsilon(D_0\cap\partial D_1\cap\cdots\cap\partial D_k) = (-)^k
\dd \chi^{(\epsilon)}_{D_k} \cdots \dd\chi^{(\epsilon)}_{D_1}
\chi^{(\epsilon)}_{D_0}\in\dRn^{(k)}\quad \forall \epsilon>0, \eqend
\reseteqn and we see that we have an example of a GDAPI as discussed
above.

{\bf Example 2:} For all $D\in\DD^{(0)}$ and $0 < \epsilon < 1$,
$\chi^{(\epsilon)}_D$ define operators $Q_D^{(\epsilon)}$ on $\cH$ \Ref{cH},
$(Q_D^{(\epsilon)}f)(x) = \chi_D^{(\epsilon)}(x)f(x)$ for all $f\in\cH$, and
these operators are in $\CCn^{(0)}$. Thus
\eq
\eta_\epsilon(Q^{(-k)}) = (-\ii)^k \ccr{\eps}{Q^{(\epsilon)}_{D_k}}\cdots
\ccr{\eps}{Q^{(\epsilon)}_{D_1}} Q^{(\epsilon)}_{D_0}
\eqend
are in $\CCn^{(k)}$ and define a GDAPI $(\Sn,\dd,\int,{^*}\Sn,\partial)$ as
discussed above.  We do not attempt here to specify $\Sn$ and just note
that it at least contains $\dRn$ as follows from our discussion above.
Especially eq.\ \Ref{result} implies that this GDAPI is a natural
noncommutative generalization of the GDAPI $(\dRn,\dd,\int,\DD,\partial)$
from Example 1.  Other examples are obtained by generating ${^*}\Sn$ with
other families of bounded, selfadjoint operators $Q_i$ with $0\leq Q_i\leq 1$
and determining $\Sn$ such that \Ref{ntr} holds.

\section{Noncommutative Descent Equations}
\subsection{Result}
{\bf Definitions:}
Given a GDA $(\CC,\dd)$, we can interpret $\CC^{(1)}$ as generalized
Yang--Mills connections and $\CC^{(0)}$ as Lie algebra of the gauge
group acting on $\CC^{(1)}$ as
$$
\CC^{(1)}\times\CC^{(0)}\to \CC^{(1)},\, (A,X)\mapsto \dd(X) +
\ccr{A}{X},
$$
and on all polynomial functions
$f:\CC^{(1)}\to\CC,\, A\to f(A)$ by Lie derivative
\eq
\cL_X f(A)\equiv \left.\f{\dd}{\dd t}f\left(A+ t\left(\dd(X) +
\ccr{A}{X}\right) \right) \right|_{t=0}
\quad\forall X\in\CC^{(0)}.
\eqend
We also define a modified Lie derivative
\eq
\label{cL}
\hat \cL_X f(A) \equiv \cL_X f(A) -\ccr{f(A)}{X}
\eqend
which obviously also obeys the Leibnitz rule, 
$\hat\cL_X(ff')=\hat\cL_X(f)f' + f\hat\cL_X(f').$

In the following we consider $k$-chains which we define as
polynomial functions
\eq
f^{k}:\underbrace{\CC^{(0)}\times\cdots\times\CC^{(0)}}_{
\mbox{{\small $k$ times}}}\times \CC^{(1)}\to \CC,\quad (X_1,\cdots,X_k,A)
\mapsto
f^{k}(X_1,\cdots,X_k;A)
\eqend
which are multilinear in the first $k$ arguments, and
antisymmetric, i.e.\ for all permutations $\pi$ of
$(1,\cdots,k)$,
\eq
f^{k}(X_{\pi(1)},\cdots,X_{\pi(k)};A) = (-)^{{\rm deg}(\pi)}
f^{k}(X_1,\cdots,X_k;A)
\eqend
where ${\rm deg}(\pi)$ is $0$ and $1$ for even and odd permutations
$\pi$, respectively. On such chains we define the operator
\eqa
\label{del}
\del f^{k-1}(X_1,\cdots,X_{k};A)=\sum_{j=1}^{k}(-)^{j+1} \cL_{X_j}
f^{k-1}(X_1,\cdots,\kill{X_j},\cdots, X_{k};A) \nonu +
\sum_{\stackrel{i,j=1}{i<j}}^{k}(-)^{i+j}
f^{k-1}(\ccr{X_i}{X_j},X_1,\cdots,\kill{X_i},\cdots,\kill{X_j},
\cdots, X_{k};A)
\eqaend
where $\kill{X_j}$ means that $X_j$ is omitted.  $\del$ is the usual 
operation acting on chains and satisfying $\del^2=0$.  We also introduce 
the operator $\hat\del$ which is defined as in eq.\ \Ref{del} but with 
$\hat \cL_{X_j}$ \Ref{cL} instead of $\cL_{X_j}$.  We write this operator 
as $\hat\delta=\delta-\iota$ with $\iota$ the operator acting on chains as
\eqa
\label{iota}
\iota f^{k-1}(X_1,\cdots,X_{k};A) = \sum_{j=1}^{k}(-)^{j+1}
\ccr{f^{k-1}(X_1,\cdots,\kill{X_j},\cdots, X_{k};A)}{X_j} .  
\eqaend

For positive, even integers $2n$, we define the $2n$--forms
\eq
\label{Ch}
Ch_{2n}(A)\equiv \left(F_A\right)^{n},
\quad F_A \equiv \dd(A) + A^2\quad
\forall A\in\CC^{(1)}\, ;
\eqend
$F_A$ is the curvature associated with the Yang--Mills connection $A$.

To write our formulas in compact form we introduce Grassmann
variables $\tet_i$ and integration $\int\dd\tet_i$
over them satisfying usual relations
\eqa
\label{Grassmann}
\car{\tet_i}{\tet_j}=\car{\tet_i}{\dd \tet_j}=
\car{\dd \tet_i}{\dd \tet_j}=\ccr{\tet_i}{\om}=\ccr{\dd \tet_i}{\om}=0
\nonu
\int\dd\tet_i\tet_j = \delta_{ij},\quad
\int\dd\tet_i=0\quad \forall i,j=0,1,2,\ldots 2n, \quad \om\in\CC.
\eqaend

For $A\in\CC^{(1)}$, $X_i\in\CC^{(0)}$, $0\leq t\leq 1$ we set
\alpheqn
\eqa
\label{nu}
\nu_{2n-k-1}^{k}(X_1,\ldots,X_{k};A,t)\equiv \int\dd\tet_{k}
\cdots \dd\tet_0
\left(F_{tA} + \tet_0 A +
\sum_{i=1}^{k} (t-1)\tet_i \dd(X_i)\right)^n \nonu
k=0,1,\ldots, n-1   \eqaend
where $F_{tA} = t\dd(A) + t^2 A^2$, and
\eqa
\label{nub}
\nu_{2n-k-1}^{k}(X_1,\ldots,X_{k};A,t)\equiv (-)^n\int\dd\tet_{k}
\cdots \dd\tet_0  
\left(\, \sum_{i=1}^k t\tet_i\dd(X_i) + \right.
\sum_{\stackrel{i,j=1}{i<j}}^k (t^2-t)\tet_i\tet_j\ccr{X_i}{X_j}
\nonu \left.
+\sum_{i=1}^k \tet_0 \tet_i X_i \right)^n,
\quad\quad k=n,n+1,\cdots 2n-1 \nonu
\eqaend
\reseteqn
(note that the  $\nu^{k}_{2n-k-1}$ for $k>n$ are actually
independent of $A$). Here and in the following we write
$\int\dd\tet_{k}\cdots\dd\tet_0$ short for
$\int\dd\tet_{k}\int\dd\tet_{k-1}\cdots\int\dd\tet_0$.

Using this we then define $k$-chains
\alpheqn
\eqa
\label{om}
\om_{2n-k-1}^{k}(X_1,\ldots,X_{k};A)\equiv \int_0^1\dd t\,
\nu_{2n-k-1}^{k}(X_1,\ldots,X_{k};A,t),\\
\tom_{2n-k-1}^{k}(X_1,\ldots,X_{k};A)\equiv \int_0^1\dd t \,t\,
\nu_{2n-k-1}^{k}(X_1,\ldots,X_{k};A,t)  \\
k=0,1,\cdots 2n-1   \nonumber
\eqaend
\reseteqn
which are easily seen to be forms of the
degrees indicated (this follows from \Ref{Grassmann}).

{\bf Theorem:}  For arbitrary graded differential algebra (GDA) $(\CC,\dd)$
and $n=1,2,\ldots$, the forms defined above obey the following generalized
descent equations,
\alpheqn
\eq
\label{desca}
Ch_{2n} =\dd(\om_{2n-1}^{0}) + \car{A}{\tom_{2n-1}^{0}}
\eqend
\eqa
\label{descb}
\del\om_{2n-k}^{k-1} + \dd(\om_{2n-k-1}^{k}) =
\iota \om_{2n-k}^{k-1} - \ccr{A}{\tom_{2n-k-1}^{k}}_\sigma
 \nonu
\sigma=(-)^k,  \quad k=1,2,\cdots n-1
\eqaend
\eq
\label{descbb}
\del\om_{n}^{n-1} + \dd(\om_{n-1}^{n}) =
\iota \om_{n}^{n-1}
\eqend
\eqa
\label{descc}
\del\om_{2n-k}^{k-1} + \dd(\om_{2n-k-1}^{k}) =
\iota \tom_{2n-k}^{k-1}  \nonu
k=n+1,\cdots 2n-1
\eqaend
\eq
\label{descd}
\del\om_{0}^{(2n-1)} = \iota\tom_{0}^{(2n-1)}.
\eqend
\reseteqn
(We recall that $\ccr{a}{b}_\sigma=ab + \sigma ba$ for $\sigma=\pm$.)

{\bf Corollary:} For all differential graded algebras with partial
integration structure (GDAPI) $(\CC,\dd,\int,{^*}\CC,\partial)$ the
forms defined above obey the following relations
\eqa
\del\int_Q \om^{k-1}_{2n-k} =
\int_{\partial Q}\om^{k}_{2n-k-1}\nonu
\forall k=1,2\cdots 2n,\quad n=1,2,\cdots, \quad Q\in{^*}\CC
\eqaend
(we set $\om_{-1}^{2n}\equiv 0$, and $\delta\int_Q\equiv\int_Q\del$).

\subsection{Proof}
We will divide the proof in three parts (a)--(c).

For the following we note that every element $\om_k\in\CC^{(n)}$
naturally defines an operator $L_{\om_k}:\CC\to\CC,
\om\mapsto L_{\om_k}(\om)\equiv \om_k \om$ (left multiplication), and obviously
$\dd L_{\om_k}= L_{\dd{\om_k}} + (-)^k L_{\om_k}\dd$.  Since the
distinction of $L_{\om_k}$ from $\om_k$ would clutter notation, we simply
write $L_{\om_k}= \om_k$.  The former relation then is written as $\dd
\om_k =
\dd(\om_k) + (-)^k \om_k\dd$.  It is therefore important in the following
to distinguish between $\dd \om$ and $\dd(\om)$ ($\om\in\CC$). Thus for
$A\in\CC^{(1)}$,   $\dd(A)=\car{\dd}{A}$, and since $\dd^2=0$ we can
write $F_A=\left(\dd +A \right)^2$. In this notation Bianchi identity
is trivial, $\ccr{\dd +A}{F_A}=0$. Similarly for $X\in\CC^{(0)}$,
$\dd(X)=\ccr{\dd}{X}$.

{\bf (a)}  We write ($\dt\equiv \partial/\partial t$)
$$
Ch_{2n}(A) = \int_{0}^1 \dd t \dt(F_{tA})^n = \int_{0}^1
\dd t\sum_{\nu=0}^{n-1}
(F_{tA})^{n-1-\nu}\dt(F_{tA})(F_{tA})^\nu
$$
(note that $F_0=0$). Since $\dt (F_{tA})=\dd(A)+2tA^2=\car{\dd +tA}{A}$
and $\dd+tA$ commutes with $F_{tA}$ (Bianchi identity), this is equivalent
to
$$
Ch_{2n}(A) = \int_{0}^1 \dd t \car{\dd +tA}{ \sum_{\nu=0}^{n-1}
(F_{tA})^{n-1-\nu} A (F_{tA})^\nu }.
$$
We note that (cf. \Ref{nu})
\eq
\nu^{0}_{2n-1}(A;t) = \int \dd\tet_0 \left(F_{tA} + \tet_0 A \right)^n
= \sum_{\nu=0}^{n-1} (F_{tA})^{n-1-\nu} A (F_{tA})^\nu,
\eqend
and with our definition \Ref{om} we obtain \Ref{desca}.

{\bf (b)} Here $k=1,2,\cdots n$.

We shall use
\eq
\cL_X(A)=\ccr{\dd+A}{X},\quad\cL_X (F_A) =\ccr{F_A}{X}.
\eqend
This implies
\eq
\label{LFt}
\cL_X(F_{tA})=\ccr{F_{tA}}{X} + (t-1)\car{\dd+tA}{\ccr{\dd}{X}}.
\eqend
(To see this, note that $F_{tA}=tF_A + (t^2-t)A^2$, thus
\nonueqa
\cL_X(F_{tA})= t\ccr{F_A}{X}+(t^2-t)\car{A}{\ccr{\dd+A}{X}} = \\
=\ccr{tF_A}{X} + \ccr{(t^2-t)A^2}{X} +(t-1)\car{tA}{\ccr{\dd}{X}}
\nonueqaend
where we used $\car{A}{\ccr{A}{X}}=\ccr{A^2}{X}$. Adding
$(t-1)\car{\dd}{\ccr{\dd}{X}}=0$ yields \Ref{LFt}.)

With Grassmann numbers $\tet_j$ as above we introduce
\eq
\label{cF}
\cA\equiv \tet_0 A + \sum_{j=1}^k (t-1)\tet_j\ccr{\dd}{X_j}, \quad
\cF_\pm \equiv F_{tA} \pm \cA
\eqend
and the operator
\eq
\Delta = \tet_0 \dt + \sum_{j=1}^k \tet_j \hat\cL_{X_j},
\eqend
and we calculate $\Delta(\cF_+)$ which we write as $(\cdot)_1+(\cdot)_2$.
Using $\dt (F_{tA})=\car{\dd +tA}{A}$,
$$
(\cdot)_1\equiv
\tet_0\dt(\cF_+) = \car{\dd +tA}{\tet_0 A} +
\sum_{j=1}^k\tet_0\tet_j\ccr{\dd}{X_j}.
$$
The relations above and the definition \Ref{cL} of $\hat\cL_X$
imply
$$
\hat\cL_X(A)=\ccr{\dd}{X},\quad
\hat\cL_X(F_{tA})=(t-1)\car{\dd+tA}{\ccr{\dd}{X}}
$$
and $\hat\cL_X(Y)=-\ccr{Y}{X}=\ccr{X}{Y}$ if $Y$ is independent of
$A$, thus
\nonueqa
(\cdot)_2
\equiv\sum_{i=1}^k\tet_i\hat\cL_{X_i}(\cF_+) = \sum_{i=1}^k\tet_i
\left(
(t-1)\car{\dd+tA}{\ccr{\dd}{X_i}} + 
\tet_0\ccr{\dd}{X_i} +
\right. \\ +\sum_{j=1}^k \left.
(t-1)\tet_j\ccr{X_i}{\ccr{\dd}{X_j}} \right) =
\car{\dd +tA}{\sum_{i=1}^k(t-1)\tet_i\ccr{\dd}{X_i}} +
\sum_{i=1}^k\tet_i\tet_0\ccr{\dd}{X_i}+\Xi
\nonueqaend
where we introduced
$
\Xi=\sum_{i,j=1}^k\tet_i\tet_j(t-1)\ccr{X_i}{\ccr{\dd}{X_j}},
$
or equivalently (using $\car{\tet_i}{\tet_j}=0$, and the Jacobi
identity and antisymmetry of $\ccr{\cdot}{\cdot}$)
\eq
\label{Xi}
\Xi=\sum_{\stackrel{i,j=1}{i<j}}^k
(t-1)\tet_i\tet_j\ccr{\dd}{\ccr{X_i}{X_j}}.
\eqend
Putting this together we see that the second terms in $(\cdot)_1$
and $(\cdot)_2$ cancel each other,
and we obtain
\eq
\label{rel}
\Delta(\cF_+) = \car{\dd +tA}{\cA} + \Xi.
\eqend

To proceed it is convenient to use another Grassmann number
$\tet\equiv \tet_{k+1}$ which satisfies $\tet\cA=-\cA\tet$.
This allows to rewrite \Ref{rel} as $\tet\Delta(\cF_+) =
\ccr{\tet(\dd+tA)}{\cA}+\tet\,\Xi$, or
equivalently (since $\ccr{\tet(\dd+tA)}{F_{tA}}=0$)
$$
\tet\Delta(\cF_+) = \ccr{\tet(\dd+tA)}{\cF_+}+\tet\,\Xi.
$$
Since $\tet\Delta\cF_+=\tet\Delta(\cF_+)+\cF_+\tet\Delta$, this
implies
\nonueqa
\tet\Delta((\cF_+)^n) =
\sum_{\nu=0}^{n-1}(\cF_+)^{n-1-\nu}\tet\Delta(\cF_+)(\cF_+)^{\nu} \\
=\sum_{\nu=0}^{n-1}(\cF_+)^{n-1-\nu}\left(\ccr{\tet(\dd+tA)}{\cF_+}
+\tet\,\Xi \right)(\cF_+)^{\nu} \\
=\ccr{\tet(\dd+tA)}{(\cF_+)^n}+
\sum_{\nu=0}^{n-1}(\cF_+)^{n-1-\nu}\tet\,\Xi(\cF_+)^{\nu}
\nonueqaend
as
$\sum_{\nu=0}^{n-1}(\cF_+)^{n-1-\nu}\ccr{\cdot}{\cF_+}(\cF_+)^{\nu}=
\ccr{\cdot}{(\cF_+)^n}$.
To get rid of $\tet$ again we use $\cF_+\tet=\tet\cF_-$
(cf.\ \Ref{cF}), thus
\eq
\label{basis}
\Delta((\cF_+)^n) = (\dd + tA)(\cF_+)^n - (\cF_-)^n(\dd +tA) +
\sum_{\nu=0}^{n-1}(\cF_-)^{n-1-\nu} \Xi(\cF_+)^\nu .
\eqend

We now apply integration
$
\int\dd\tet_k\cdots \dd\tet_0
$
to this equation. In the following calculation
we keep in mind that we can always forget about terms containing
higher powers than $1$ of at least one of the $\tet_i$'s. Thus,
\nonueqa
\int\dd\tet_k\cdots \dd\tet_0 \Delta((\cF_+)^n)=
\int\dd\tet_k\cdots \dd\tet_0 \tet_0\dt \left(F_{tA}+ \sum_{j=1}^k
(t-1)\tet_j\ccr{\dd}{X_j}\right)^n \\
+ \sum_{i=1}^k\hat\cL_{X_i} \int\dd\tet_k\cdots \dd\tet_0 \tet_i
\left(F_{tA}+ \tet_0 A+ \sum_{\stackrel{j=1}{j\neq i}}^k
(t-1)\tet_j\ccr{\dd}{X_j}\right)^n.
\nonueqaend
Using
$$
\int\dd\tet_k\cdots \dd\tet_0 \tet_j =
(-)^j\int\dd\tet_k\cdots \kill{\dd\tet_j}\cdots \dd\tet_0\quad 
j=0,1\cdots k,
$$
the definition \Ref{nu} of $\nu_{2n-k}^{k-1}$ and introducing the 
$k$--chains 
\eq
\beta_{2n-k}^k(X_1,\ldots,X_k;A,t) = 
\int\dd\tet_k\cdots \dd\tet_1  \left(F_{tA}+ \sum_{j=1}^k
(t-1)\tet_j\ccr{\dd}{X_j}\right)^n , 
\eqend
we can write this
\nonueqa
\label{this}
\int\dd\tet_k\cdots \dd\tet_0\Delta((\cF_+)^n)=
\dt\beta_{2n-k}^k(X_1,\ldots,X_k;A,t)
\\+
\sum_{j=1}^k(-)^j\hat\cL_{X_j}\nu_{2n-k}^{k-1}(X_1,\ldots,
\kill{X_j}, \ldots, X_k;A,t).
\nonueqaend
Moreover, since $\int\dd\tet_k\cdots \dd\tet_0 (\cF_\pm)^n =
(\pm)^{k+1}\nu_{2n-k-1}^{k}(X_1,\ldots,X_k;A,t)$ (cf.\ \Ref{nu} and
\Ref{cF}),
we have
\nonueqa
\int\dd\tet_k\cdots \dd\tet_0
\left( (\dd+tA)(\cF_+)^n
- (\cF_-)^n(\dd+tA)\right)  \\ =
\ccr{\dd+tA}{\nu_{2n-k-1}^{k}(X_1,\ldots,X_k;A,t)}_\sigma,
\quad \sigma=(-)^k . 
\nonueqaend
Note that this formula remains also true for $k=n$ if we set 
$\nu^n_{n-1}$ to zero.  
Finally (we use $\cF_-\tet_i=\tet_i\cF_+$; cf.\ \Ref{Xi}),
\nonueqa
\int\dd\tet_k\cdots \dd\tet_0 \sum_{\nu=0}^{n-1}(\cF_-)^{n-1-\nu}
\Xi(\cF_+)^\nu   \\ =
\sum_{\stackrel{i,j}{i<j}}\sum_{\nu=0}^{n-1}
\int\dd\tet_k\cdots \dd\tet_0\tet_i(\cF_+)^{n-1-\nu}
(t-1)\tet_j\ccr{\dd}{\ccr{X_i}{X_j}}(\cF_+)^\nu  \\ =
\sum_{\stackrel{i,j}{i<j}}
\int\dd\tet_k\cdots \dd\tet_0\tet_i \left(F_{tA} + \tet_0 A +
(t-1)\tet_j\ccr{\dd}{\ccr{X_i}{X_j}} +
\sum_{\stackrel{\ell=1}{\ell\neq i,j }}^{k}
(t-1)\tet_\ell\ccr{\dd}{X_\ell} \right)^n \\
=
\sum_{\stackrel{i,j}{i<j}}(-)^{i+j}\nu^{k-1}_{2n-k}
(\ccr{X_i}{X_j},X_1,\cdots,\kill{X_i},\cdots,\kill{X_j},
\cdots, X_{k};A,t)
\nonueqaend
where we used
$$
\int\dd\tet_k\cdots \dd\tet_0\tet_i = (-)^{i+j}
\int\dd\tet_k\cdots\kill{\dd\tet_i}\cdots\kill{\dd\tet_j}
\cdots  \dd\tet_1  \dd\tet_j \dd\tet_0 .
$$
Using the definitions \Ref{del} and \Ref{iota} we can therefore write
$\int\dd\tet_k\cdots\dd\tet_0$\Ref{basis} as
\eq
\label{that}
\hdel\nu_{2n-k}^{k-1}
+ \ccr{\dd+tA}{\nu_{2n-k-1}^{k}}_\sigma
 =
\dt\beta_{2n-k}^k,
\quad \sigma=(-)^{k}
\eqend
where we suppressed the common argument $(X_1,\ldots,X_k;A,t)$.  Our 
descent equations for $k=1,\ldots,n-1$ are obtained by applying integration 
$\int_0^1\dd t$ to \Ref{that}.  We first note that for $k=1,\ldots,n-1$, 
$\beta_{2n-k}^k(X_1,\ldots,X_k;A,t)$ is zero for $t=1$ (trivialy) and $t=0$ 
(since $F_0=0$), thus $\int_0^1\dd t\dt\beta_{2n-k}^{k}=0$.  Thus with the 
definitions \Ref{om} and $\hdel=\del-\iota$,
\eq
\del \om_{2n-k}^{k-1} - \iota\om_{2n-k}^{k-1} +
\ccr{\dd}{\om_{2n-k-1}^{k}}_\sigma +
\ccr{A}{\tom_{2n-k-1}^{k}}_\sigma =0, \quad \sigma=(-)^k,
\quad k=1,2,\ldots n-1
\eqend
equivalent to \Ref{descb}. For $k=n$, eq.\ \Ref{that} remains true 
with $\nu^{n}_{n-1}$ set to zero, thus
\eq
\label{above}
\del\om^{n-1}_n - \iota \om^{n-1}_n = N^{n}_n
\eqend
where $N^{n}_n\equiv \int_0^1\dd t\dt \beta_{n}^{n}$ is nonzero,
\eq
\label{Om}
N^{n}_n(X_1,\ldots,X_n) = (-)^{n+1}
\int\dd\tet_n\cdots\dd\tet_1\left(
\sum_{i=1}^n\tet_i\dd(X_i) \right)^n .
\eqend
To complete the proof of \Ref{descbb}, we will show below 
that $N^{n}_n$ equals  $-\dd(\om_{n-1}^{n})$.

{\bf (c)} Here $k=n,n+1,\cdots 2n-1$.

We introduce
\eq
\cV = \sum_{i=1}^k\tet_i X_i,\quad \ccF_\pm= \pm t\dd(\cV) + (t^2-t)\cV^2
+ \tet_0 \cV \eqend
and the operator \eq
\dDel = \tet_0\dt + t\hI,\quad \hI = \sum_{i=1}^k \tet_i
\ccr{X_i}{\cdot}. \eqend
We observe
$$
\cV^2 = \half\sum_{i,j=1}^k\tet_i\tet_j\ccr{X_i}{X_j}=
\sum_{\stackrel{i,j=1}{i<j}}^k \tet_i\tet_j\ccr{X_i}{X_j}
$$
implying $\hI(\cV)= 2\cV^2$. Similarly,
$$
\hI(\dd(\cV)) = \half\sum_{i,j=1}^k\tet_i\tet_j\left(
\ccr{X_i}{\dd(X_j)} - \ccr{X_j}{\dd(X_i)}\right) = \dd(\cV^2),  $$
and
\nonueqa
\hI(\cV^2)= \sum_{\ell=1}^k
\sum_{\stackrel{i,j=1}{i<j}}^k
\tet_\ell \tet_i\tet_j\ccr{X_\ell}{\ccr{X_i}{X_j}}
\\
=\mbox{$\f{1}{6}$}\sum_{i,j,\ell=1}^k
\tet_\ell\tet_i\tet_j\left(\ccr{X_\ell}{\ccr{X_i}{X_j}}
+ \ccr{X_i}{\ccr{X_j}{X_\ell}} + \ccr{X_j}{\ccr{X_\ell}{X_i}}
\right) = 0
\nonueqaend
where we used $\tet_\ell\tet_i\tet_j=\tet_i\tet_j\tet_\ell$ etc.\ and
the Jacobi identity for $\ccr{\cdot}{\cdot}$. Thus $$
t\hI(\ccF_+) =  t^2\dd(\cV^2) -2t\tet_0 \cV^2 $$
(note that $\hI\tet_0=-\tet_0\hI$). Obviously $\tet_0\dt(\ccF_+) =
\tet_0\dd(\cV) + (2t-1)\tet_0\cV^2$. Combining the last two equations
we obtain
\eq
\dDel(\ccF_+) = \ccr{\dd}{\ccF_+} + \cXi,\quad
\cXi = t\dd(\cV^2) -(t^2-t)\hI(\cV^2) -\tet_0\cV^2
\eqend
where we added terms $\dd(t\dd(\cV)) = -2(t^2-t)\hI(\cV^2)=0$ and used 
$\dd(\ccF_+)= \ccr{\dd}{\ccF_+}$. Similarly as in (b) above this implies
\eq
\label{aha}
\dDel((\ccF_+)^n) = \dd(\ccF_+)^n - (\ccF_-)^n\dd +  
\sum_{\nu=0}^{n-1}(\ccF_-)^{n-1-\nu}\cXi (\ccF_+)^\nu \: . 
\eqend
Noting that our definition \Ref{nub} can be written as
$\nu^{k}_{2n-k-1}=(-)^n\int\dd\tet_k\cdots\dd\tet_0(\ccF_+)^n$,
we now apply $(-)^n\int\dd\tet_k\cdots\dd\tet_0$ to this equation.
The following calculation is similar to the one above in (b),
\nonueqa
(-)^n\int\dd\tet_k\cdots\dd\tet_0 \dDel((\cF_+^n)) =
\dt\tilde\beta^{k}_{2n-k}(X_1,\cdots,X_k;t) \\
+ \sum_{i=1}^k (-)^{i} t\ccr{X_i}{\nu^{k-1}_{2n-k}(X_1,\ldots,
\kill{X_i}, \ldots, X_k;t)}
\nonueqaend
where we introduced
\eq
\label{tnu}
\tilde\beta^{k}_{2n-k}(X_1,\cdots,X_k;t) =
(-)^n\int\dd\tet_k\cdots\dd\tet_1\left(t\dd(\cV)+(t^2-t)\cV^2\right)^n
\eqend
and used the definition \Ref{nub}.  Note that this remains true 
also for $k=n$ if we set $\nu^{n-1}_n$ to zero.  Similarly as in 
(b), $$ (-)^n\int\dd\tet_k\cdots\dd\tet_0 \left( \dd(\ccF_+)^n - 
(\ccF_-)^n\dd
\right)  = 
\ccr{\dd}{\nu^{k}_{2n-k-1}(X_1,\ldots,X_k;t)}_{\sigma}, \quad 
\sigma=(-)^{k} . 
$$
To calculate the last term we write
\nonueqa
\cXi = \sum_{\stackrel{i,j=1}{i<j}}^k \tet_i \left(
t\tet_j\dd(\ccr{X_i}{X_j})  +
\sum_{\stackrel{\ell=1}{\ell\neq i,j}}^{k}(t^2-t) 
\tet_j\tet_\ell \ccr{\ccr{X_i}{X_j}}{X_\ell}
+ \tet_0\tet_j\ccr{X_i}{X_j}
\right) ,
\nonueqaend
and similarly as in (b),
\nonueqa
(-)^n\int\dd\tet_k\cdots \dd\tet_0 \sum_{\nu=0}^{n-1}(\ccF_-)^{n-1-\nu}
\cXi(\ccF_+)^\nu   \\ =
\sum_{\stackrel{i,j}{i<j}}(-)^{i+j}\nu^{k-1}_{2n-k}
(\ccr{X_i}{X_j},X_1,\cdots,\kill{X_i},\cdots,\kill{X_j},
\cdots, X_{k};t) . \nonueqaend
Recalling the definitions \Ref{del} and \Ref{iota} we can therefore
write $(-)^n\int\dd\tet_k\cdots \dd\tet_0$\Ref{aha} as
\eqa
\label{abc}
\del\nu^{k-1}_{2n-k} + \ccr{\dd}{\nu^{k}_{2n-k-1}}_\sigma =
\dt\tilde\beta^{k}_{2n-k} + \iota t \nu^{k-1}_{2n-k},
\quad \sigma=(-)^k 
\eqaend
where we suppressed the common argument $(X_1,\ldots,X_k;t)$ (we used 
that $\cL_{X_j}$ acting on anything independent of $A$ gives zero).
Again the descent equations are obtained by integrating
$\int_0^1\dd t$ this equation. Since for $k>n$,
$\tilde\beta^{k}_{2n-k}$ is zero for $t=0$ and
$t=1$ we get
\eq
\del\om^{k-1}_{2n-k} + \dd(\om^{k}_{2n-k-1}) =
\iota\tom^{k-1}_{2n-k-1}
\eqend
proving \Ref{descc} for $k=n+1,\ldots,2n-1$. Obviously this equation
remains also true for $k=2n$ if we set $\dd(\om^{2n}_{-1})=0$. This
implies \Ref{descd}.

For $k=n$, \Ref{abc} remains true if we set $\nu^{n-1}_{n}$ to zero. Moreover,
it is easy to see that $\int_0^1\dd t \dt \tilde\beta^{n}_{n} = -
N^{n}_n$ as defined in \Ref{Om} (cf.\ \Ref{tnu}), thus
\eq
\dd(\om^{n}_{n-1}) = -  N^{n}_n . 
\eqend
Combining this with equation \Ref{above}
gives \Ref{descbb} which completes the proof of the theorem.

The corollary then follows trivially from our definitions, especially the
defining relations \Ref{GDAPI} of integrations on GDAPI.

\section{Final Remarks} Given a GDA $(\CC,\dd)$ one can consider a
$\N_0$--graded vector space $C=\bigoplus_{k=0}^\infty C^{(k)}$ of
polynomial maps
$$
c^{k}:\underbrace{\CC^{(0)}\times\cdots\times\CC^{(0)}}_{
\mbox{{\small $k$ times}}}\times \CC^{(1)}\to \C,\quad (X_1,\cdots,X_k,A)
\mapsto
c^{k}(X_1,\cdots,X_k;A)
$$
which are multilinear and antisymmetric in the first $k$ arguments,
with the operator $\del:C^{(k)}\to C^{(k+1)}$ defined as in \Ref{del}. 
Since $\del^2=0$, $(C,\del)$ is a graded differential
complex. A map $c^k\in C^{(k)}$ is called a $k$--cocycle if $\del
c^k =0$, and a $k$--cocycle $c^k$ is nontrivial if it is not a
$k$--boundary i.e.\ not of the form $\del b^{k-1}$ with $b^{k-1}\in
C^{(k-1)}$.
Anomalies in quantum gauge theories are nontrivial cocycles, e.g.\ the
axial anomalies and Schwinger terms are nontrivial 1-- and 2--cocycles,
respectively \cite{Jackiw}.

The result of this paper in Section 3 provides an explicit construction of
$k$--cocycles
\eq
\label{ex}
c^k_{2n-k-1} = \int \omega^k_{2n-k-1}
\eqend
for arbitrary GDA $(\CC,\dd)$ with integration $\int$.  The cocycle
property is a trivial consequence of our generalized descent equations and
basis properties of integration.  Whether or not $c^k_{2n-k-1}$ is
nontrivial depends on the integration $\int$ and the precise definition
of $C^{(k-1)}$ --- in most cases it will be
trivial or even zero.  Well-known examples for \Ref{ex} being nontrivial is
for the de Rham GDA $(\dR{2n-k-1},\dd)$ with integration
$\int_{\R^{2n-k-1}}\trc$, or more generally, the de Rham GDA $(\dR{d},\dd)$
, $d$ big enough, with integration $\int_{M}\trc$ over a $(2n-k-1)$
dimensional submanifold $M$ of $\R^{d}$.  According to our discussion in
Section 2, these examples immediately generalize to universal Rham GDA (=
Example 2; the hats in the following are to indicate that we are considering
universal de Rham forms). We conjecture that all the
universal cocycles $\hat c^k_{2n-k-1}$ for $(\CCs{2n-k-1},\hd,\hint)$
are nontrivial since this is known to be true for $k=2$ and $n=2$ and
$n=3$. Indeed,
$\hat c^2_1$ is proportional to the Lundberg cocycle
\cite{Lund} known to be nontrivial, and $\hat c^2_3$ is proportional to the
Mickelsson--Rajeev cocycle which is also nontrivial \cite{MR}.

To have one explicit example and also demonstrate how formulas can be
written without using Grassmann numbers, we finally give a more explicit
expression for the universal Schwinger terms generalizing the
Mickelsson--Rajeev cocycle to higher dimensions,
\eqa
\label{final}
\hat c^2_{2n-3}(\hat X_1,\hat X_2;\hat A) = \f{3}{c_{2n-3}}
\sum_{\stackrel{i_\nu\geq 0}{\mbox{\tiny{$\sum$}}_\nu i_\nu = n-3}}
(\ii)^2 \int_0^1\dd t
(t-1)^2 \nonu\times
\TraC{
(\hat F_{tA})^{i_1} \hat A
(\hat F_{tA})^{i_2}\ccr{\eps}{\hat X_1}(\hat F_{tA})^{i_3}\ccr{\eps}{\hat
X_2}(\hat F_{tA})^{i_4}
- (\hat X_1\leftrightarrow \hat X_2) }\quad (n\geq 3) 
\eqaend
where $\hat F_{tA} = \ii\car{\eps}{\hat A}t +
\hat A^2 t^2 $
(we used cyclicity under conditional Hilbert space trace; for
notation see Section 2); the $\hat X_i$ and $\hat A$ here are bounded
operators on a Hilbert space $\cH$ such that $\ccr{\eps}{\hat X_i}$ and
$\ccr{\eps}{\hat A}$ both are in the Schatten class $\cB_{2n-2}$ and
$\car{\eps}{\hat A}\in \cB_{n-1}$.  Restricting to operators $\hat A=
(F-\eps)$ where $F=F^{-1}=F^*$ is a grading operator
(as usually done in this context \cite{L}), we have
$\hat F_{tA}=(t^2-t)(F-\eps)^2$, and \Ref{final} should be proportional
to the universal Schwinger terms given in \cite{FT}.


\bc{\bf Acknowledgments}\ec
I am grateful to D. Burghelea, A. Carey, H. Grosse, G. Landi, and
especially G. Kelnhofer and J. Mickelsson for useful discussions.  This
paper was written at the the Erwin Schr\"odinger International Institute in
Vienna. I would like to thank the ESI for hospitality and P.  Marmo and
P. Michor for inviting me there. I thank C. Ekstrand for carefully 
reading the manuscript. 


\end{document}